\documentclass[a4paper,12pt]{elsarticle}

\usepackage[top=0.5in,bottom=0.75in,left=0.6in,right=0.6in]{geometry}
\usepackage{graphicx}

\usepackage[figurename=Fig.]{caption}
\usepackage[tablename=Tab.]{caption}
\usepackage{color}
\usepackage{float}
\usepackage{amssymb}
\usepackage{amsmath}
\usepackage{bbm}
\usepackage{mathrsfs}
\usepackage{textcomp}

\usepackage{cases}
\usepackage{upgreek}
\usepackage{makecell}
\usepackage{pdflscape}
\usepackage{subfigure}
\usepackage{pdfpages}
\usepackage{lineno}
\usepackage{makeidx}
\usepackage{bm}
\usepackage{xstring}
\modulolinenumbers[1]
\usepackage{epstopdf}
\usepackage{lipsum}
\usepackage{slashed}
\usepackage{multicol}
\usepackage{setspace}
\usepackage{booktabs}
\usepackage[table]{xcolor}
\usepackage{multirow}
\usepackage{braket}

\usepackage[separate-uncertainty]{siunitx}

\newcommand{\eref}[1]{Eq.\,(\ref{#1})}
\newcommand{\w}{\omega}
\newcommand{\vo}{\vec{o}\@ifnextchar{^}{\,}{}}
\def\up{\mathrm}
\def\degree{^\circ}

\renewcommand{\thetable}{\Roman{table}}

\newcommand{\er}[2]{^{-#1}_{+#2}}

\graphicspath{{PDF/}}

\definecolor{myblue}{RGB}{65,111,166}
\definecolor{myred}{RGB}{168,66,63}
\definecolor{mygreen}{RGB}{134,164,74}
\definecolor{mypurple}{RGB}{110,84,141}
\definecolor{myindigo}{RGB}{61,150,174}
\definecolor{myorange}{RGB}{218,129,55}
\definecolor{mylightblue}{RGB}{142,165,203}

\makeatletter
\def\ps@pprintTitle{%
  \let\@oddhead\@empty
  \let\@evenhead\@empty
  \let\@oddfoot\@empty
  \let\@evenfoot\@oddfoot
}
\makeatother



\usepackage{hyperref}
\hypersetup{
    colorlinks=true,
    linkcolor=red,
    citecolor=red,
    urlcolor=cyan,
}

\begin{document}


\begin{frontmatter}

\title{Mixing angle and decay constants of $J^P=1^+$ heavy-light mesons}

\author[1]{Qiang Li\corref{corauthor}} \ead{liruo@nwpu.edu.cn}
\author[2]{Tianhong Wang}\ead{thwang@hit.edu.cn}
\author[2]{Yue Jiang}\ead{jiangure@hit.edu.cn}
\author[3]{Guo-Li Wang \corref{corauthor}}\ead{gl\_wang@hit.edu.cn}
\author[4,5,6]{Chao-Hsi Chang}\ead{zhangzx@itp.ac.cn}
\cortext[corauthor]{Corresponding author}

\address[1]{Department of Applied Physics,  Northwestern Polytechnical University, Xi'an 710129, P. R. China}
\address[2]{Department of Physics, Harbin Institute of Technology, Harbin 150001, China}
\address[3]{Department of Physics, Hebei University, Baoding 071002, China}
\address[4]{Institute of Theoretical Physics, Chinese Academy of Sciences, P.O. Box 2735, Beijing 100080, China}
\address[5]{CCAST (World Laboratory), P.O. Box 8730, Beijing 100080, China}
\address[6]{School of Physical Sciences, University of Chinese Academy of Sciences, Beijing 100049, China}

\begin{abstract}
The mass spectra, mixing angle and decay constants of the $J^P=1^+$ heavy-light mesons are systematically studied within the framework of the Bethe-Salpeter equation (BSE). The full $1^+$ Salpeter wave function is given for the first time. The mixing between the $1^{+-}$ and $1^{++}$ in the $1^+$ heavy-light systems are automatically determined by the dynamics in the equation without any man-made mixing. The results indicate that in a rigorous study there exists the phenomenon of mixing angle inversion or mass inversion within $1^{+}$ heavy-light doublet, which is sensitive to the $s$-quark mass for the charmed mesons and $u$- or $d$-quark masses for the bottomed mesons. This inversion phenomenon can answer the question of why we have confused mixing angles in the literature and partly explain the lower mass of $D_{s1}(2460)$ compared to that of $D_{s1}(2536)$. The decay constants are also presented and can behave as a good quantity to distinguish the $1^+$ doublet in heavy-light mesons. This study indicates that the light-quark mass may play an important role in deciding the mass order, mixing angle, and decay constant relation between the $\ket{j_l=\frac{3}{2}}$ and $\ket{j_l=\frac{1}{2}}$ heavy-light mesons. 
%
%
\end{abstract}

\end{frontmatter}


\section{Introduction}
Generally, all the physical mesons have definite $J^P$ spin parity or $J^{PC}$ for quarkonia. The spin $S$ and orbital angular momentum $L$ are no longer the good quantum numbers in the relativistic situations, and usually the physical states are not located in the definite $^{2S+1}\!L_J$ states. These situations become obvious in the $1^+$ and $1^-$ mesons; for the $1^-$ states, the $2^3\!S_1$-$1^3\!D_1$ mixing is needed to fit the experimental measurements for both quarknia\,\cite{Rosner2005,Barnes2005} and heavy-light mesons\,\cite{Zhong2010,LiD2010,LiD2011,Chen2011,GM2014,Song2015,Chen2015}, while for the $1^+$ states, we always have to make the $^1P_1$-$^3P_1$ mixing fit the physical states\,\cite{Cahn2003,Rosner1986,GK1991}.
So, to describe the bound states more effectively and appropriately, one should focus on the $J^{P(C)}$, which are always the good quantum numbers.
In the previous literature, the unnatural parity $1^+$ heavy-light mesons were usually studied by two methods, one is the heavy quark effective theory~(HQET)\,\cite{Rosner1986,GK1991}, and another makes a man-made mixing between the $^1\!P_1$ and $^3\!P_1$ states.
For the former one, which works in the approximation $m_Q\!\to\!+\infty$, and it does not hold well when the light-quark mass is comparable with the heavy quark, such as in the $(c\bar s)$ and $(b\bar c)$ systems. While for the latter one, the mixing angle is always difficult to decide and usually treated as a free parameter. Neither of the two methods to deal with the unnatural parity states is satisfactory.

On the other hand, the mass relation between the two $1^+$ states is also a problem. The mass of the broad state $D_{1}(2430)$ is little heavier than that of the narrow state $D_{1}(2420)$, while compared with the narrow state $D_{s1}(2536)$, the broad state $D_{s1}(2460)$ has a much lower mass. In the relativized Godfrey-Isgur\,(GI) model\,\cite{GI1985}, the masses of the $1^+$ $(c\bar s)$ doublet are predicted to be 2.55 and 2.56 GeV\,\cite{GK1991,GM2016}, which correspond to the experimental $D_{s1}(2535)$ and $D_{s1}(2460)$ respectively in the traditional quark model. This is the famous low-mass puzzle, which means the mass of $D_{s1}(2460)$ is much lower than the quark model predictions\,\cite{GI1985,GM2016,Pierro2001,Song2015A}. A more detailed review on this low-mass puzzle can be found in Ref.\,\cite{Chen2016}.
 The coupled channel effects (CCEs) \cite{Beveren2003,Beveren2004} have been used to answer the low-mass question of $D_{s1}(2460)$. But we want to explore what the mass relation would be between the two heavy-light $1^+$ states, when the CCEs can be ignored or only make small contribution. A long time ago, Schnitzer first noted that, according to the spin-orbit interaction between quarks, there may be inverted mass relations between $\ket{\frac{1}{2}}$ and $\ket{\frac{3}{2}}$ multiplets\,\cite{Schnitzer1978,Isgur1998}; this is not right for the $0^+$ and $2^+$ states, but we want to know if this would happen to the two $1^+$ states.

\begin{table}[!h]
\caption{The discovered $J^P=1^+$ heavy-light mesons from the experimental information of the	Particle Data Group (PDG)\,\cite{PDG2018} and the corresponding predictions of the GI model\,\cite{GI1985,GK1991,GM2016}. The mass and width are in units of MeV.}\label{Tab-Exp}
\vspace{0.2em}\centering
\begin{tabular}{ lllll }
\toprule[2pt]
{Resonances}   			& Mass$_{\up{Exp.}}$     				  &Mass$_\up{GI}$&Width    					&Decay     		 \\
\midrule[1.5pt]
$D_1(2420)^0$   			& $2421.4\pm0.6$       &2.46& $27.4\pm2.5$ 	 		&$D^{*+}\pi^{-}$\\
$D_1(2420)^\pm$ 		& $2423.2\pm2.4$       &2.46& $25\pm6$ 	 	 		&$D^{*0}\pi^{0}$ \\
$D_1(2430)^0$   			& $2427\pm36$    	  &2.47& $384^{+107}_{-75}\pm75$ 	&$D^{*+}\pi^{-}$ \\
$D_{s1}(2460)$  			& $2459.5\pm0.6$       &2.56& $<3.5$ 					&$D_s^*\pi^0,~D_s\gamma$\\
$D_{s1}(2536)$  			& $2535.1\pm0.1$       &2.55& $0.92\pm0.05$ 			&$D^{*}K$\\
$B_1(5721)^0$		     	& $5727.7\pm2.0$       &5.78& $30.1\pm3.8$ 			&$B^{*+}\pi^-$\\
$B_1(5721)^+$ 		  	& $5725.1\pm2.0$       &5.78& $29.1\pm5.6$ 			&$B^{*}\pi^+$\\
$B_{s1}(5830)^0$			& $5828.7\pm0.4$	  &5.86& $0.5\pm0.4$ 				&$B^{*}K$\\
\bottomrule[2pt]
\end{tabular}
\end{table}

In fact, from the view of experiments, the $1^+$ heavy-light mesons have not been well established\,\cite{PDG2018}. In \autoref{Tab-Exp} the current known mesons with $J^P=1^+$ are listed.  In the nonrelativistic description, the $J^P=1^+$ doublet is generally considered as the mixtures of the  $^1P_1$ and $^3P_1$ states,
\begin{equation} \label{E-R}
\begin{pmatrix}\ket{P_l}\\ \ket{P_h} \end{pmatrix}
=R(\theta)\begin{pmatrix}|^1P_1\rangle \\ |^3P_1\rangle \end{pmatrix}
= \begin{bmatrix}\cos\theta & \sin\theta \\ -\sin\theta& \cos\theta\end{bmatrix} \begin{pmatrix}|^1P_1\rangle \\ |^3P_1\rangle \end{pmatrix},
\end{equation}
where $\ket{P_l}$ and $\ket{P_h}$ denote the lower- and higher-mass state, respectively; and $R(\theta)$ is the defined mixing matrix with angle $\theta$; and $^1P_1$ and $^3P_1$ correspond to the $J^{PC}=1^{+-}$ and $1^{++}$, respectively. For neutral charmed mesons $D_1(2420)$ and $D_1(2430)$, the mixing angle $\theta({D_1})\simeq 35.3\degree$\,\cite{Cahn2003} is determined in the heavy-quark limit. In the traditional quark models, the analogy $1^+$ charm-strange doublet is also considered as the mixtures of $^1P_1$ and $^3P_1$ states.
However, in order to fit the experimental data, this time, one has to use the mixing angle $\theta({D_{s1}})=-54.7\degree$\,\cite{Barnes2005,Song2015,WangZH2018}. The different choices of mixing angles in charm and charm-strange systems caused ambiguities in the previous literature.
In this work, we will try to show and explain the different choices by the full $1^+$ Salpeter wave functions. In the bottomed systems, the $1^+$ states $B_1(5721)^0$, $B_1(5721)^+$, and $B_{s1}(5830)^0$ are discovered in experiments, while their orthogonal partners and the two $1^+$ $B_{c1}^{(\prime)}$ states are still missing\,\cite{PDG2018}. We will also explore the mixing angle and mass spectra, and especially discuss whether the mixing angle inversions exist in the $J^P=1^+$ bottomed systems.

The decay constant is another physical quantity we are interested, which appear in many weakly decay processes and are quite important in extracting some fundamental quantities, such as the Cabibbo–Kobayashi–Maskawa (CKM) matrix elements. Also under the factorization assumption\,\cite{Fakirov1978,Cabibbo1978,Bauer1987}, the decay constants play a key role in calculating the nonleptonic decays. So besides the mixing angle and mass spectra, we will also calculate the decay constants of the $1^+$ heavy-light mesons, which could behave as a cross-check on our analysis. 

In this work, we will directly construct the Salpeter wave function for $J^P=1^+$ states without using any man-made mixing angle. By solving the corresponding Salpeter wave functions, we could naturally obtain the mixing angle of the $1^+$ heavy-light mesons.
This work is studied within the framework of the instantaneous Bethe-Salpeter\,(BS) methods\,\cite{SB1951,Salpeter1952},
which have been widely used and have achieved good performance in the strong decays of heavy mesons\,\cite{Chang2005,WangT2013,WangT2017}, hadronic transition\,\cite{WangT2013B,Ju2015,LiQ2016}, decay constants calculations, and annihilation rates\,\cite{WangGL2006, WangGL2009, WangT2013A}. This manuscript is organized as follows. In \autoref{Sec-2},  first, we construct the BS wave function of the $1^+$ states and then calculate the mixing angle and decay constants. In \autoref{Sec-3}, we present the numerical results and discussions of the mixing angle and decay constants. Finally, we give a short summary of this work.

\section{Theoretic calculations}\label{Sec-2}
In this section, first, we give a brief review of the instantaneous BS methods; then we present the formalism of mixing angle and decay constants together with BS wave function of $J^P=1^+$ states.

\subsection{\rm Brief review on the instantaneous BS methods}
The Bethe-Salpeter equation of the meson in momentum space reads\,\cite{SB1951}
\begin{gather}
\Gamma(P,q)=\int \frac{\up d^4 k}{(2\pi)^4}iK(k-q)[ S(k_1)\Gamma(P,k) S(-k_2)] ,
\end{gather}
where $\Gamma(P,q)$ is the BS vertex; $P$ is the total momentum of the meson;  and $S(k_1)$ and $S(k_2)$ are the Dirac propagators of the quark and antiquark, respectively. The internal momenta $q$ and $k$ are defined as,
\[q=\alpha_2p_1-\alpha_1p_2,~~k=\alpha_2k_1-\alpha_1k_2; \]
$\alpha_i \equiv \frac{m_i}{m_1+m_2}~(i=1,2)$,  where $m_{1(2)}$ denotes the constituent mass of the quark (antiquark), and $p_1(k_1)$ and $p_2(k_2)$ are the corresponding momenta. The BS wave function of the meson is then defined as
\begin{gather} \label{E-BSwave}
 \psi(P,q)\equiv S(p_1)\Gamma(P,k) S(-p_2).
\end{gather}

As usual, in this work, the specific interaction kernel we use is the Coulomb-like potential  plus the unquenched scalar confinement one.
In the instantaneous approximation, the interaction kernel does not depend on the time component of $s=(k-q)$. Then, the QCD-inspired interaction kernel used in this work is 
\begin{equation}
K(s)\simeq  K(\vec s\,)=\left[V_\up{G}(\vec s\,)+V_0\right]\gamma_\mu\otimes\gamma^\mu+V_\up{C}(\vec s\,),
\end{equation}
where the potential in the Coulomb gauge behaves as\,\cite{Chao1992,DingYB1993,DingYB1995,Kim2004}
\begin{equation}
V_\up{G}(\vec s\,)= -\frac{4}{3} \frac{4\pi \alpha_s(\vec s\,)}{\vec s\,^2+a_1^2},~~~V_\up{C}(\vec s\,)=(2\pi)^3 \delta^3(\vec s\,) \frac{\lambda}{a_2}- \frac{8\pi \lambda}{(\vec s\,^2+a_2^2)^2},
\end{equation}
where $\frac{4}{3}$ is the color factor; $a_{1(2)}$ is introduced to avoid the divergence in small momentum transfer zone; and the kernel describing the confinement effects is introduced phenomenologically, which is characterized by the the string constant $\lambda$ and the factor $a_2$. The potential used here originates from the famous Cornell potential\,\cite{Eichten1978,Eichten1980}, namely, the one-gluon exchange Coulomb-type potential at short distance and a linear growth confinement one at long distance. To incorporate the color screening effects\,\cite{Laermann1986,Born1989} in the linear confinement potential, $V_\up{C}$ is modified and taken as the aforementioned form. $V_0$ is a free constant fixed by fitting the data. The strong coupling constant $\alpha_s$ has the form,
\begin{equation}\notag
\alpha_s(\vec s\,)=\frac{12\pi}{(33-2N_f)}\frac{1}{\ln\left(a+\frac{\vec s\,^2}{\Lambda^2_{\up{QCD}}}\right)},
\end{equation}
where $\Lambda_\up{QCD}$ is the scale of the strong interaction, $N_f$ is the active flavor number, and $a=e$ is a constant. In this work, we will only consider the time component $(\mu=0)$ of the vector kernel, for the spatial components $(\mu=1,2,3)$ are always suppressed by a factor $\frac{v}{c}$ in the heavy-light meson systems.

With the instantaneous kernel, we can introduce the three-dimensional BS wave function (also called the Salpeter wave function) $\varphi(q_\perp) \equiv  i\int \frac{\text{d}q_P}{2\pi}\psi(q)$, where $q_P=\frac{q\cdot P}{M}$ corresponds to $q^0$ in the rest frame of $P$. Then we can express the BSE as a three-dimensional integration equation,
\begin{gather}
\Gamma(q_\perp)=\int \frac{\text{d}^3k_\perp}{(2\pi)^3}K(k_\perp -q_\perp)\varphi(k_\perp),
\end{gather}
where $q_\perp=q-q_P\frac{P}{M}$; and $\Gamma(q_\perp)$ is the three-dimensional BS vertex.
$S(p_1)$ and $S(-p_2)$ are the propagators for the quark and antiquark, respectively. To perform the integration over $q_P$, we decompose the propagators as
\begin{equation}\label{E-Propagator}
\begin{aligned}
S(+p_1)&=\frac{i\Lambda_1^+}{q_P+\alpha_1M-\omega_1+i\epsilon}+\frac{\si{i}\Lambda_1^-}{q_P+\alpha_1M+\omega_1-{i}\epsilon},\\
S(-p_2)&=\frac{i\Lambda_2^+}{q_P-\alpha_2M+\omega_2-i\epsilon}+\frac{\si{i}\Lambda_2^-}{q_P+\alpha_2M-\omega_2+{i}\epsilon},
\end{aligned}
\end{equation}
where $\omega_i=\sqrt{m_i^2-p_{i\perp}^2}$, and the projection operators are defined as
\[
\Lambda^{\pm}_1=\frac{1}{2}\left[ 1\pm \hat{H}(p_{1\perp})\right] \gamma^0,~~\Lambda^{\pm}_2=\frac{1}{2}\gamma^0 \left[ 1\mp \hat{H}(p_{2\perp})\right],
\]
where $\hat H(p_{i\perp}) \equiv \frac{1}{\w_i}(p^\alpha_{i\perp}\gamma_\alpha+m_i)\gamma^0$ are the usual Dirac Hamilton divided by $\w_i$.

Performing the contour integration over $q_P$ on both sides of Eq.~(\ref{E-BSwave}),  the BSE is reduced to the following four coupled three-dimensional Salpeter equations\,\cite{Salpeter1952}
\begin{equation}
\begin{gathered}
(M-\omega_1-\omega_2)\varphi^{++}(q_\perp)=+\Lambda_1^+(q_\perp) \Gamma(q_\perp)\Lambda^+_2(q_\perp) ,\\
(M+\omega_1+\omega_2)\varphi^{--}(q_\perp)=-\Lambda_1^-(q_\perp) \Gamma(q_\perp)\Lambda^-_2(q_\perp) ,\\
\varphi^{+-}(q_\perp)=\varphi^{-+}(q_\perp)=0,
\end{gathered}
\end{equation}
where $\varphi^{\pm\pm}$ are defined as $\varphi^{\pm\pm}\equiv\Lambda_1^\pm(q_\perp)\frac{\slashed P}{M}\varphi(q_\perp)\frac{\slashed P}{M}\Lambda_2^\pm(q_\perp) $; $\varphi^{++}$ and $\varphi^{--}$ are called the positive and negative energy wave functions, respectively; and in the weak bound states usually we have $\varphi^{++}\gg \varphi^{--}$; and it can be easily checked that $\varphi=\varphi^{++}+\varphi^{-+}+\varphi^{+-}+\varphi^{--}$.
Note that the Salpeter equations are, in fact, two eigenvalue equations and two constraint conditions. The bound state mass $M$ behaves as the eigenvalue.
The normalization condition for Salpeter equation reads
\begin{gather}
\int \frac{\text{d}^3q_\perp}{(2\pi)^3} \up{Tr}\left[\overline\varphi^{++}\frac{\slashed P}{M}\varphi^{++}\frac{\slashed P}{M}-\overline\varphi^{--}\frac{\slashed P}{M}\varphi^{--}\frac{\slashed P}{M}\right]=2M.
\end{gather}

The Salpeter equations can also be rewritten as the compact Shr\" odinger type,
\begin{align}\label{E-SE}
M\varphi(P,q_\perp)
&=(\w_1+\w_2)\hat H(p_{1\perp})\varphi(q_\perp) +\frac{1}{2} \left[ \hat H(p_{1\perp})W(q_\perp) - W(q_\perp) \hat H(p_{2\perp})\right],
\end{align}
with the constraint condition,
\begin{equation} \label{E-BS-constraint}
\hat  H(p_{1\perp}) \varphi(p_\perp) + \varphi(p_\perp)\hat  H(p_{2\perp}) =0,
\end{equation}
where $W(p_\perp)\equiv\gamma^0\Gamma(q_\perp)\gamma_0$ denotes the potential energy part.
The normalization condition is now expressed as
\begin{equation} \label{E-SE-norm}
\int \frac{\up d^3 \vec q}{(2\pi)^3}  \up{Tr}\,\varphi^{\dagger}(P,q_\perp)  \hat H(p_{1\perp})\varphi(P,q_\perp) =2M.
\end{equation}

\subsection{\rm Salpeter wave function of the $1^+$ states}\label{Sec-2-2}
To solve the above Salpeter equation, we have to construct the form of the wave function according to the different spin-parity $J^{P}$ and appropriate Dirac structures. The Salpeter wave function for $J^P=1^+$ states will be given in this subsection. It is the first time that the $1^+$ Salpeter wave functions are obtained without using the artificial mixing. The mixing between $1^{+-}$ and $1^{++}$ for the $1^+$ doublet will be determined naturally by the dynamics of the BSE without using any free mixing angle.

The general form of the $J^P=1^+$ states Salpeter wave functions can be constructed as 
\begin{gather}\label{E-1+wave}
\varphi_{1^{+}}= \frac{q_\perp \!\cdot\! \xi}{|\vec q\,|} \left (f_1  + f_2  \frac{\slashed P}{M} +  f_3 \frac{\slashed q_\perp}{|\vec q\,|} + f_4 \frac{\slashed P \slashed q_\perp}{M|\vec q\,|}   \right)\gamma^5 + i\frac{\epsilon_{\mu P q_\perp \xi}}{M|\vec q\,|}\gamma^\mu \left(h_1+ h_2 \frac{\slashed P}{M}+  h_3 \frac{\slashed q_\perp}{|\vec q\,|} +  h_4  \frac{\slashed P \slashed q_\perp}{M|\vec q\,|} \right),
\end{gather}
where the radial wave functions $f_{i}(|\vec q\,|)$ and $h_i(|\vec q\,|)\, (i=1,\cdots,4)$ are  explicitly dependent on $|\vec q\,|$; $\epsilon_{\mu P q_\perp \xi}=\epsilon_{\mu \nu \alpha \beta }{P^\nu q^\alpha_\perp \xi^\beta}$ and $\epsilon_{\mu \nu \alpha \beta}$ is the totally antisymmetric Levi-Civita tensor, and $\xi$ is the polarization vector of the bound state and fulfills $P\cdot \xi=0,~\sum \xi_\mu^{(r)}\xi_\nu^{(r)}=\frac{P_\mu P_\nu}{M^2}-g_{\mu \nu}$.
Moreover, the constraint condition, Eq.\,(\ref{E-BS-constraint}), can further reduce the undetermined radial wave functions to 4, namely
\begin{equation}
\begin{aligned}
f_3  &  =-\frac{|\vec q\,|(\omega_1-\omega_2)}{m_1\omega_2+m_2\omega_1}f_1,  &\quad  f_4  &  =-\frac{|\vec q\,|(\omega_1+\omega_2)}{m_1\omega_2+m_2\omega_1}f_2;  \\
h_3 &  = +\frac{|\vec q\,|(\w_1-\omega_2)}{m_1\omega_2+m_2\omega_1}h_1,         &\quad h_4 &  = +\frac{|\vec q\,|(\w_1+\omega_2)}{m_1\omega_2+m_2\omega_1}h_2.
 \end{aligned}
\end{equation}
Notice that $f_{3(4)}$ and $h_{3(4)}$ are suppressed by a factor of $|\vec q\,|$.
Now there only exist four independent radial wave functions $f_1,~f_2,~h_1$, and $h_2$.
Inserting this wave function into Eq.\,(\ref{E-SE-norm}), we obtain the normalization condition as
\begin{gather}
\braket{f_1f_2}-2\braket{h_1h_2}=1,~~~\braket{x_1x_2}\equiv \int \frac{\up{d}^3\vec q}{(2\pi)^3} \frac{8\w_1\w_2}{3M(m_1\w_2+m_2\w_1)}(x_1x_2).
\end{gather}
where we defined the abbreviation $\braket{x_1x_2}$ to denote the normalization integral.

It can be checked that, the first part of $\varphi_{1^+}$, consisting of $f_1,~f_2,~f_3$, and $f_4$, has the spin parity $J^{PC}=1^{+-}$, while the second part, consisting of $h_1,~h_2,~h_3$, and $h_4$, has $J^{PC}=1^{++}$. The $J^P=1^+$ Salpeter wave function can also be expanded in terms of the spherical harmonics $Y_l^m$, and then we can find that it also contains the $S$- and $D$-wave components besides the dominant $P$-wave (see appendix\ref{App}). Then, we decompose the $1^+$ Salpeter wave function \eref{E-1+wave} into two parts according to \eref{E-R},
\begin{gather}
\varphi_{l}=+\cos\theta\varphi_{1^{+-}}+\sin\theta \varphi_{1^{++}},  \label{E-phi-l}\\
\varphi_{h}=-\sin\theta\varphi_{1^{+-}}+\cos\theta \varphi_{1^{++}}, \label{E-phi-h}
\end{gather}
where $\varphi_{1^{+-}}$ and $\varphi_{1^{++}}$ are the normalized Salpeter wave functions for $1^{+-}$ and $1^{++}$ states, respectively. 
Then the mixing angle $\theta$ can be obtained from the integral of the low-mass wave function $\varphi_l$ as,
\begin{equation} \label{E-Theta-l}
\cos^2\theta=\braket{f_1f_2}_l,~~~~\sin^2\theta=-2\braket{h_1h_2}_l.
\end{equation}
Of course, the mixing angle can also be calculated from the integral of $\varphi_h$ as,
\begin{equation}\label{E-Theta-h}
\cos^2\theta=-2\braket{h_1h_2}_h,~~~~\sin^2\theta=\braket{f_1f_2}_h,
\end{equation}
which would give exactly the same mixing angle as that from \eref{E-Theta-l}. Since an overall minus sign can be absorbed by the redefinition of $\varphi_{l(h)}$, we can constraint the mixing angle to a range of $-90^\circ$ to $90^\circ$.
The relative sign of $\theta$ can be determined by the relative sign between $f_i$ and $h_i$. For example, if the signs of $(f_1,h_1)$ for $\varphi_l(c\bar u)$ are $(+,-)$, and $(+,+)$ for $\varphi_l(c\bar s)$, we can conclude that their mixing angles should  differ by a minus sign. Also notice that $\varphi_h(\theta)=\varphi_l(\theta+90^\circ)$, namely, the two states in the $J^P=1^+$ doublet are orthogonal, and we can use the form of \eref{E-phi-l} to express the general $J^P=1^+$ Salpeter wave function, in which the low- and high-mass states are denoted by the mixing angle $\theta$ and $(\theta+90^\circ)$ respectively. More about the mixing angle will be discussed in the next section.

\begin{figure}[htpb]
\vspace{0.5em}
\centering
\subfigure[BS wave function for $1P_l (c\bar u)$.]{\includegraphics[width=0.40\textwidth]{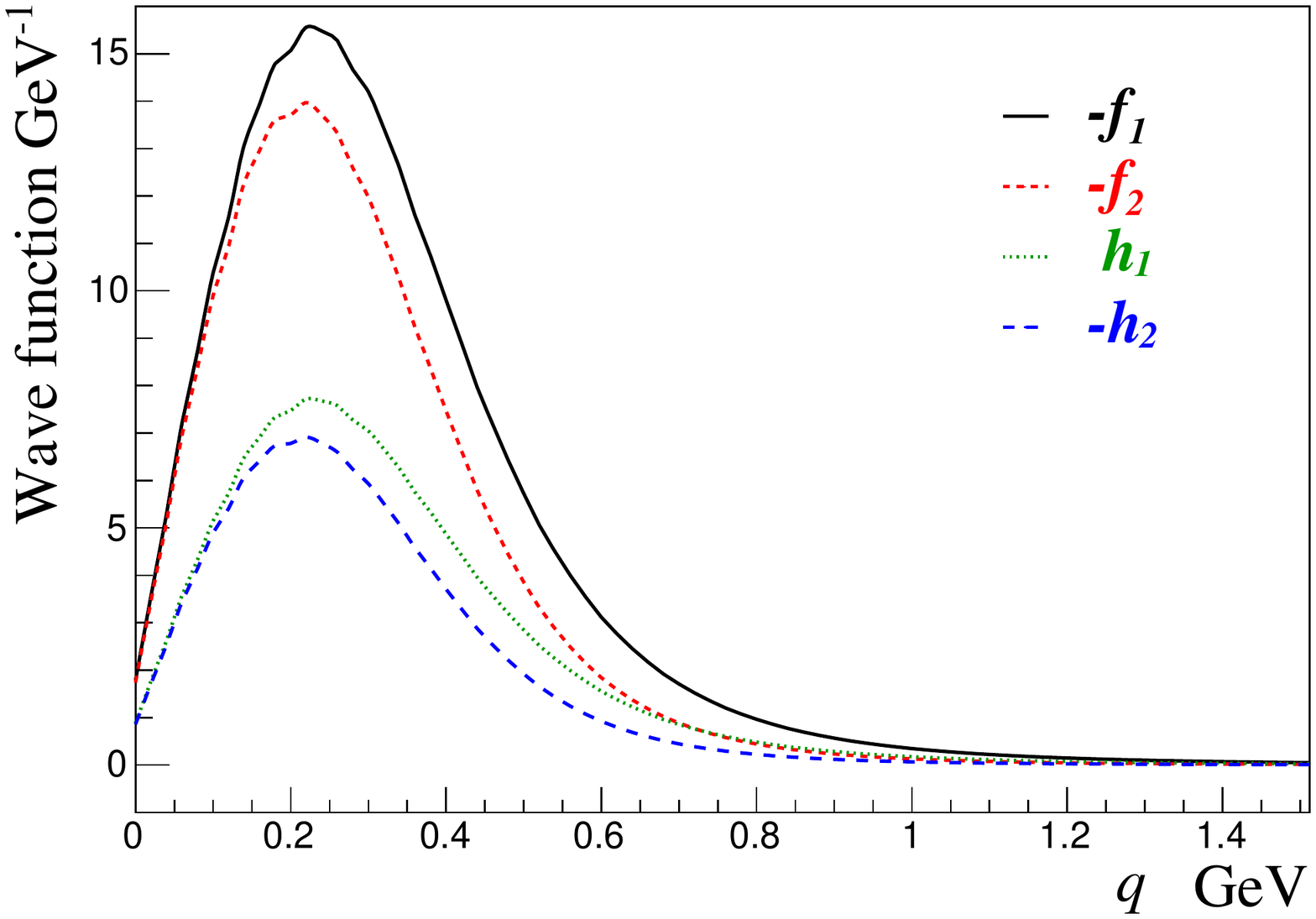} \label{Fig-cu-n1}}
\subfigure[BS wave function for $1P_h (c\bar u)$.]{\includegraphics[width=0.40\textwidth]{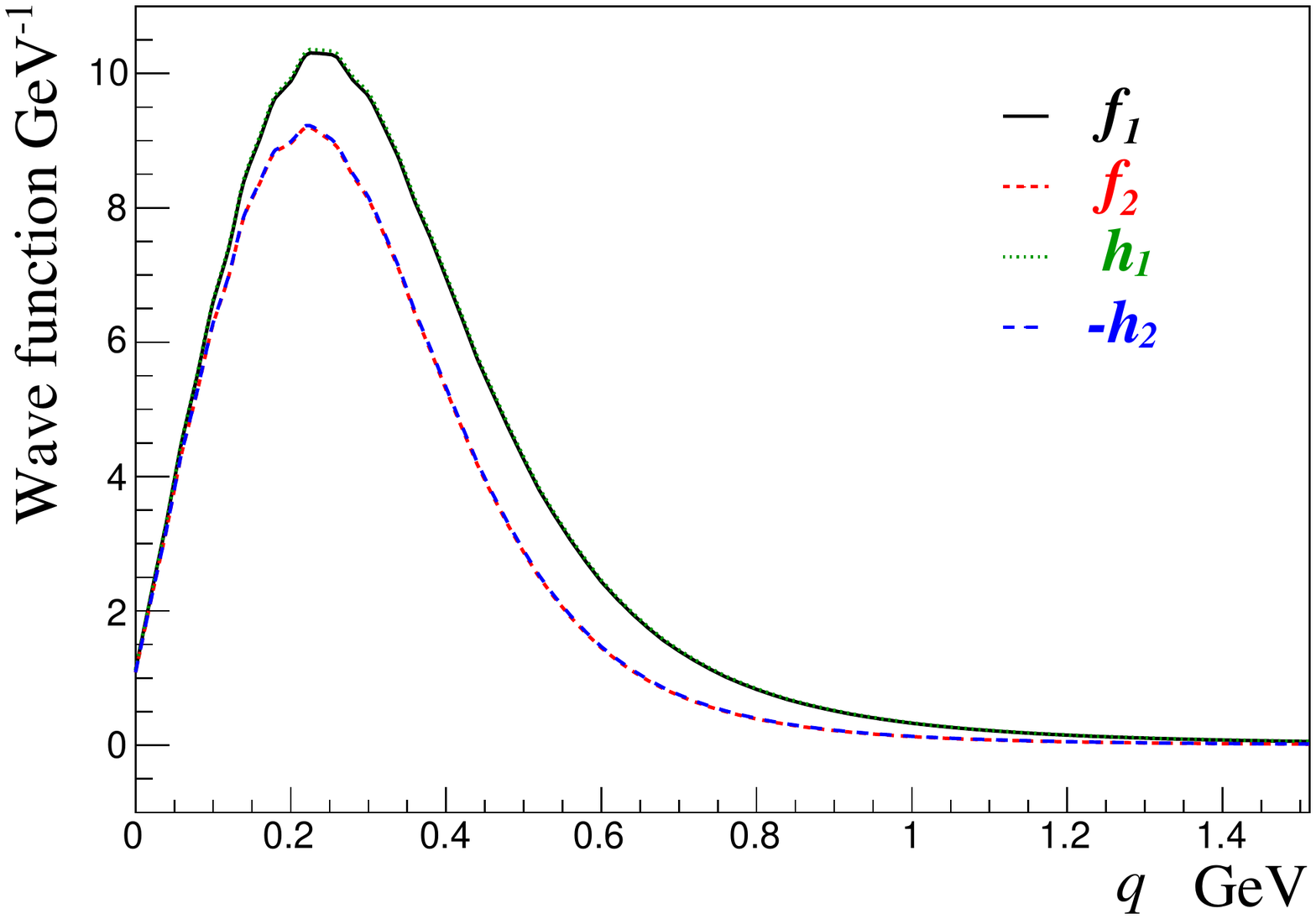} \label{Fig-cu-n2}}\\
\subfigure[BS wave function for $2P_l (c\bar u)$.]{\includegraphics[width=0.40\textwidth]{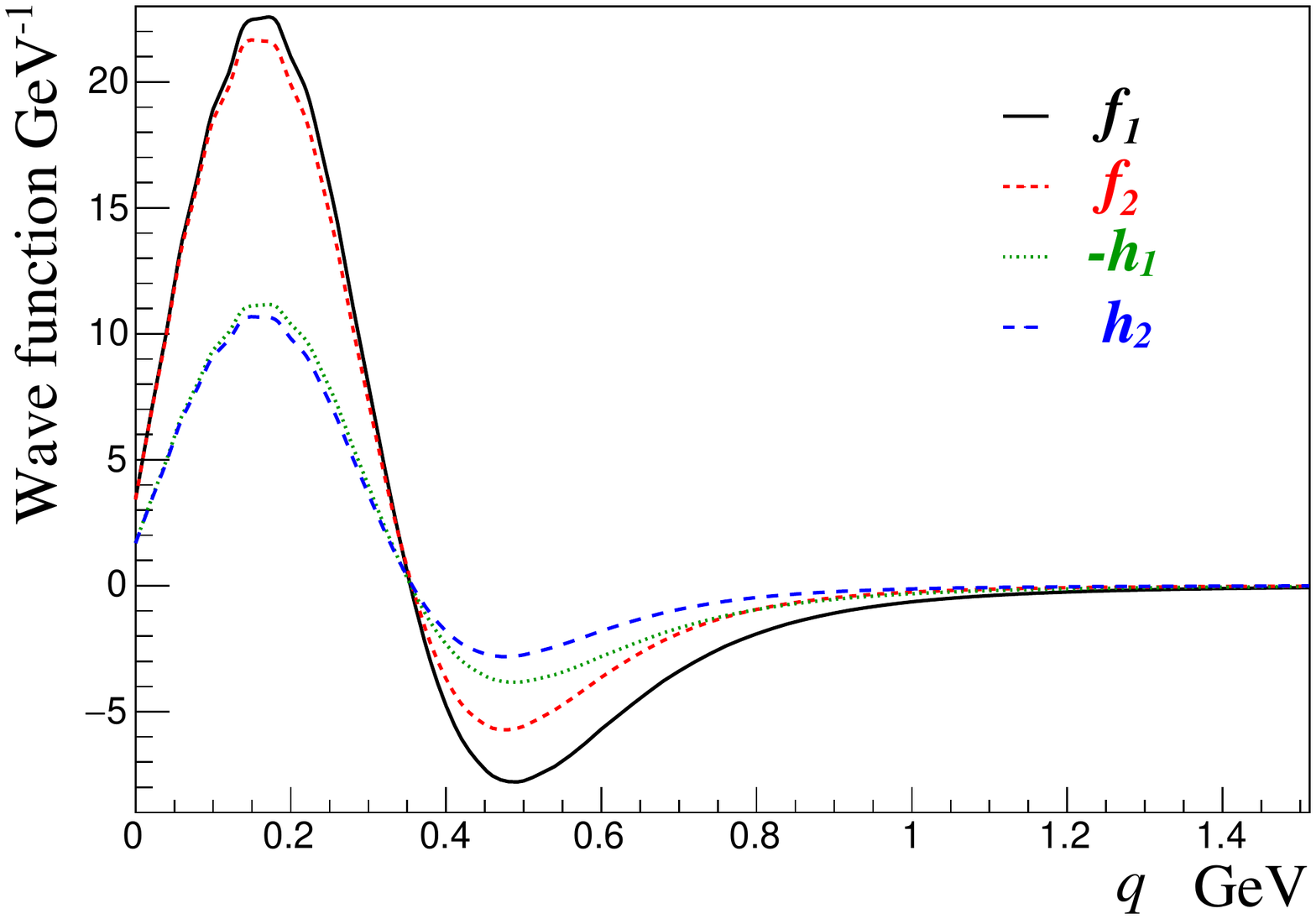} \label{Fig-cu-n3}}
\subfigure[BS wave function for $2P_h (c\bar u)$.]{\includegraphics[width=0.40\textwidth]{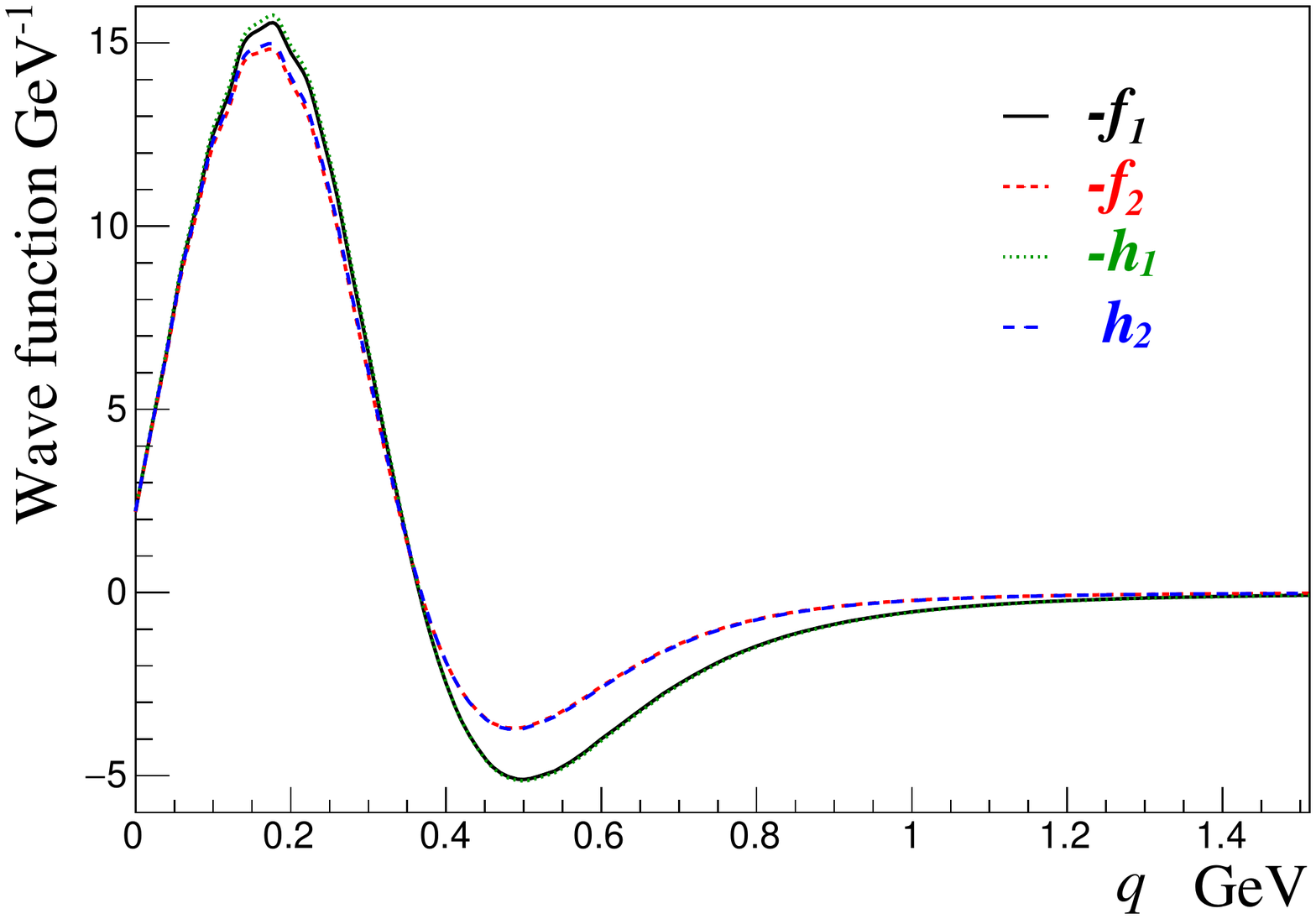} \label{Fig-cu-n4}}\\
\caption{BS wave function for $1^+$ state $D_1$ mesons.}\label{Fig-wave}
\vspace{0.5em}
\end{figure}

By solving the BS equation (the detailed procedures on solving the full Salpeter equation can be found in our previous work\,\cite{Kim2004,WangT2013B,LiQ2016,WangT2017A}), we obtain the numerical results including two sets of solutions. The wave functions share the same structure, but take different radial values; see \autoref{Fig-wave}.
In \autoref{Fig-cu-n1} and \autoref{Fig-cu-n3}, the $1^+$ $(c\bar u)$ radial wave functions of low-mass states $|nP_l\rangle$ with the radial quantum number $n=1,~2$ are shown, while the results of its corresponding partners, namely, the high-mass states $|1P_h\rangle$ and $|2P_h\rangle$ are displayed in \autoref{Fig-cu-n2} and \autoref{Fig-cu-n4}. Notice that the figures show $f_1\simeq h_1$ and $f_2\simeq-h_2$ for the ${nP_h}$ $(c\bar u)$. Then we can calculate that $\tan^2\theta\simeq\frac{1}{2}$  from \eref{E-Theta-h}; namely, the mixing angles are about $35.3^\circ$. The different structures of radial wave functions of two $J^P=1^+$ states will lead to different physics, for example, the decay constants.

Before moving on, we first discuss the nonrelativistic mixing angle predicted in the heavy-quark limit,  which could behave as a simple check for our results. In the heavy-quark limit, the total angular momentum $j_l$ of the light quark becomes the good quantum number. Then, it is more convenient to describe the heavy-light mesons in the $|J,j_l\rangle$ basis, which is related to the $|J,S\rangle$ basis by\,\cite{Cahn2003}
\begin{gather}\label{E-JjJS}
\begin{pmatrix} \ket{\frac{3}{2}} \\ \ket{\frac{1}{2}} \end{pmatrix}=
R(\theta_{H})
\begin{pmatrix} |^1P_1\rangle \\ |^3P_1\rangle \end{pmatrix}=
\frac{1}{\sqrt{3}}\begin{bmatrix}\sqrt{2} & 1\\ -1 & \sqrt{2}\end{bmatrix} \begin{pmatrix} |^1P_1\rangle \\ |^3P_1\rangle \end{pmatrix},
\end{gather}
where $\theta_H=\arctan{\sqrt{1/2}}=35.3\degree$ denotes the ideal mixing angle in heavy quark limit.
Combining Eqs.\,(\ref{E-R}) and (\ref{E-JjJS}), we can conclude that in heavy-quark limit $\theta=35.3\degree$ if the state $|\frac{3}{2}\rangle$ is the lower-mass one; while $\theta=\theta_H-90=-54.7\degree$ if the state $|\frac{3}{2}\rangle$ is the higher-mass one. It should be pointed out that the two different mixing angles arise from our mixing convention defined in Eq.\,(\ref{E-R}), in which we always put the lower-mass one upside. Apart from this, they are totally equivalent, just as stated in Ref.\,\cite{Matsuki2010}.  So, if our methods could correctly reflect the character of the heavy-light mesons, we should obtain the mixing angle $\theta$ close to the $\theta_{H}$ or $(90^\circ-\theta_{H})$.

On the other hand, from Eqs.\,(\ref{E-R}) and (\ref{E-JjJS}), the states $|P_l\rangle$ and $|P_h\rangle$ can also be expressed in the heavy-quark limit basis $|J,j_l\rangle$ as
\begin{gather}\label{E-UnTheta}
\begin{pmatrix} |P_l\,\rangle\\|P_h\rangle \end{pmatrix}=R( \theta_{H})
\begin{pmatrix} \ket{\frac{3}{2}}\\ \ket{\frac{1}{2}} \end{pmatrix},
\end{gather}
where $\theta_{H}=\theta-35.3\degree$. Usually, if above the corresponding strong decay threshold, the $|j_l=\frac{3}{2}\rangle$ state corresponds to the narrow state since it could only decay by the $D$-wave, while the $|j_l=\frac{1}{2}\rangle$ state corresponds to the broad one for it could decay by the $S$-wave. The $D_1(2420)$ and $D_1(2430)$ are just exactly coincident with the analysis. In this work, among the  $1^+$ doublet, we will always use $|nP\rangle$ to denote the $|\frac{3}{2}\rangle$ dominant state, while $|nP'\rangle$  will denote the $|\frac{1}{2}\rangle$ dominant one. In the heavy-quark limit basis, usually, one should obtain the mixing angle $\theta$ close to $0^\circ$ or $-90^\circ$.

\subsection{\rm Decay constants}
The decay constant for the $J^P=1^+$ meson is defined as
\begin{gather}
f_{1^+} M \xi^\mu \equiv \langle 0| \bar q \Gamma^\mu Q |M,\xi \rangle,
\end{gather}
where the abbreviation $\Gamma^\mu\equiv\gamma^\mu(1-\gamma^5)$ is used, and $Q$ and $\bar q$ denote the heavy-quark and light-antiquark fields, respectively. According to the Mandelstam formalism\,\cite{Mandelstam1955}, the transition matrix element can be expressed by the Salpeter wave function as,
\begin{gather}
\langle 0| \bar q_1 \Gamma^\mu q_2 |M,\xi \rangle = -\sqrt{N_c}\int \frac{\up{d}^3 q_\perp}{(2\pi)^3}\up{Tr}[\varphi(q_\perp)\Gamma^\mu]=\frac{4 \sqrt{N_c}}{3} \xi^\mu \int \frac{\up{d}^3 q_\perp}{(2\pi)^3}(f_3+2h_4),
\end{gather}
where $N_c=3$ denotes the number of colors.
Then the decay constant can be expressed by Salpeter wave function as
\begin{gather}
f_{1^+} = \frac{4 \sqrt{N_c}}{3M}  \int \frac{\up{d}^3 \vec q}{(2\pi)^3}(f_3+2h_4).
\end{gather}
From above expression, we can see that decay constant is sensitive to the relative sign of $f_3$ and $h_4$, namely, the sign of the mixing angle $\theta$.

\section{Numerical results and discussions}\label{Sec-3}
First, we specify the model parameters used in in this work.
The potential model parameters we use in this work read
\begin{align*}
a&=e=2.7183,    	 &a_1&=0.060~\si{GeV}, &\lambda&=0.125~\si{GeV}^2,   &\Lambda_\text{QCD}&=0.252~\si{GeV}, &a_2&=0.040~\si{GeV}.
\end{align*}
The constituent quark masses we use are $m_u=0.305~\si{GeV}$, $m_d=0.311~\si{GeV}$, $m_s=0.5~\si{GeV}$, $m_c=1.72~\si{GeV}$, and $m_b=4.96~\si{GeV}$. The free parameter $V_0$ is fixed by fitting the mass eigenvalue to experimental values.  Besides, the retardation effects are considered as a perturbation term and incorporated by making the replacement $\vec s\,^2\to \vec s\,^2-(s^0)^2$ in the interaction kernel, where $s^0$ is further expressed by its on-shell value by assuming the quarks\,(anti-quarks) are on their mass shells\,\cite{Qiao1996,Qiao1999,Ebert2000A}.

\begin{table}[h!]
\caption{Mass spectrum and decay constants of $1^+$ heavy-light mesons in\si{MeV}. The mixing angles are presented in units of degrees. $\theta_{nH}=\theta_{nP}-35.3\degree$, where $\theta_{nH}$ is under basis $|\frac{3}{2}\rangle$ and $|\frac{1}{2}\rangle$, while $\theta_{nP}$ is under basis $|^1P_1\rangle$ and $|^3P_1\rangle$ with $n$ denoting the radial quantum number.  }\label{Tab-M-Dc}
\vspace{0.2em}\centering
\begin{tabular}{ crrrrrrr }
\toprule[2pt]
$Q\bar q$ 		&$c\bar u$			&$c\bar d$		&$c\bar s$				&$b\bar u$		&$b\bar d$		&$b\bar s$		&$b\bar c$	\\
$V_0$     		&485 				&485   		&249					&857			&857			&710			&181\\
\midrule[1.5pt]
    $M_{1l}$ 	& $2421\er{96}{95}$  			& $2433\er{96}{94}$  		& $2531\er{85}{85}$  				& $5714\er{216}{215}$  		& $5720\er{216}{215}$  		& $5803\er{219}{217}$  		& $6815\er{218}{218}$  \\
    $M_{1h}$ 	& $2431\er{93}{92}$ 			& $2441\er{93}{92}$  		& $2535\er{85}{84}$  				& $5721\er{219}{217}$  		& $5728\er{219}{217}$  		& $5829\er{220}{218}$  		& $6830\er{217}{217}$  \\
    $M_{2l}$ 	& $2863\er{88}{88}$  			& $2873\er{88}{88}$  		& $2936\er{86}{86}$  				& $6214\er{207}{205}$  		& $6221\er{207}{205}$  		& $6305\er{205}{203}$  		& $7168\er{217}{217}$  \\
    $M_{2h}$ 	& $2878\er{88}{88}$  			& $2888\er{88}{88}$  		& $2941\er{87}{86}$  				& $6222\er{205}{203}$  		& $6228\er{205}{204}$  		& $6307\er{206}{204}$  		& $7174\er{217}{217}$  \\
    $M_{3l}$ 	& $3139\er{90}{90}$  			& $3149\er{90}{90}$  		& $3196\er{88}{88}$  				& $6522\er{199}{198}$  		& $6539\er{199}{198}$  		& $6604\er{200}{200}$  		& $7415\er{218}{218}$  \\
    $M_{3h}$ 	& $3149\er{90}{90}$  			& $3159\er{90}{90}$  		& $3200\er{89}{88}$  				& $6526\er{198}{198}$  		& $6533\er{198}{198}$  		& $6604\er{200}{200}$  		& $7419\er{218}{218}$  \\
\midrule[1.5pt]
    $f_{1l}$ 	& $56.6\er{5.2}{8.7}$  			& $57.7\er{6.3}{10.8}$  		& $267.7\er{9.0}{8.9}$ 	& $265.9\er{225.2}{8.9}$ 		& $266.6\er{9.0}{8.8}$ 		& $286.1\er{7.5}{7.1}$ 		& $227.0\er{12.8}{13.4}$ \\
    $f_{1h}$ 	& $266.8\er{8.7}{8.6}$ 			& $266.3\er{8.8}{8.9}$ 		& $54.9\er{47.2}{5.9}$  	& $20.3\er{16.4}{239.1}$  		& $21.0\er{13.8}{5.0}$  		& $33.4\er{2.4}{2.5}$  		& $57.0\er{2.3}{2.3}$  \\
    $f_{2l}$ 	& $59.9\er{3.6}{3.6}$  			& $60.5\er{4.4}{4.6}$  		& $81.5\er{6.0}{8.6}$  	& $31.5\er{3.0}{4.1}$  		& $32.0\er{3.2}{4.5}$  		& $239.1\er{165.4}{6.5}$ 		& $201.4\er{7.7}{7.6}$ \\
    $f_{2h}$ 	& $222.4\er{7.8}{8.0}$ 			& $221.8\er{7.5}{7.8}$ 		& $212.8\er{7.5}{7.5}$ 	& $240.3\er{6.3}{6.4}$ 		& $240.2\er{6.3}{6.3}$ 		& $16.9\er{16.2}{210.9}$  		& $52.1\er{2.0}{1.9}$  \\
    $f_{3l}$ 	& $59.0\er{3.0}{3.0}$  			& $59.5\er{3.9}{3.9}$  		& $78.2\er{4.7}{6.1}$  	& $33.8\er{3.1}{3.9}$  		& $34.3\er{3.2}{4.2}$  		& $222.8\er{138.2}{5.9}$ 		& $189.7\er{6.5}{6.5}$\\
    $f_{3h}$ 	& $200.1\er{7.0}{7.1}$ 			& $199.8\er{6.7}{6.9}$ 		& $194.8\er{6.8}{6.8}$ 	& $221.4\er{5.7}{5.7}$ 		& $221.3\er{5.7}{5.7}$ 		& $9.6\er{9.6}{197.6}$  		& $49.3\er{1.8}{1.7}$  \\
\midrule[1.5pt]
    $\theta_{1P}$ & $35.1\er{0.4}{1.6}$  		& $35.1\er{0.5}{2.1}$  		& $-60.4\er{10.0}{1.4}$ 		& $-55.4\er{5.3}{114.4}$ 		& $-55.4\er{10.1}{0.2}$ 		& $-55.3\er{0.1}{0.1}$ 		& $-58.0\er{0.4}{0.4}$ \\
    $\theta_{2P}$ & $34.9\er{0.2}{0.3}$  		& $34.9\er{0.3}{0.3}$  		& $36.2\er{1.1}{2.1}$  		& $35.9\er{0.3}{0.6}$  		& $35.9\er{0.3}{0.7}$  		& $-59.7\er{26.5}{114.5}$   	& $-58.8\er{0.5}{0.4}$ \\
    $\theta_{3P}$ & $35.0\er{0.3}{0.4}$  		& $35.0\er{0.3}{0.4}$  		& $36.2\er{1.1}{1.7}$ 		& $36.2\er{0.4}{0.7}$  		& $36.3\er{0.4}{0.8}$  		& $-62.0\er{24.6}{128.1}$ 		& $-59.3\er{0.5}{0.5}$ \\
\midrule[1.5pt]
    $\theta_{1H}$ & $-0.2$  		     & $-0.2$  		& $84.3$ 					& $89.3$ 			& $89.3$ 			& $89.4$ 		& $86.7$ \\
    $\theta_{2H}$ & $-0.4$  		     & $-0.4$  		& $0.9$  				     & $0.6$  		     & $0.6$  		     & $85.0$   	& $85.9$ \\
    $\theta_{3H}$ & $-0.3$  		     & $-0.3$  		& $0.9$ 				     & $0.9$  		     & $1.0$  		     & $88.7$ 		& $85.4$ \\
\bottomrule[2pt]
\end{tabular}
\end{table}

The obtained mass spectra, decay constants, and mixing angles are presented in \autoref{Tab-M-Dc}, in which we use the symbols $\theta_{nP}$ and $\theta_{nH}$ to denote the mixing angles defined in \eref{E-R} and \eref{E-UnTheta}, respectively in order to indicate the different radially excited states.
We can see clearly that there exist the $J^P=1^+$ doublet, two states with close mass and the same radial quantum number. The predicted masses of two $J^P=1^+$ $(c\bar u)$ are consistent with experimental data, while since we did not consider the effect of CCEs, the theoretical mass for $D_{s1}(2460)$ is still about $70$ MeV higher than experimental data.

The mixing angles $\theta_{1P}$ for $(c\bar u)$ and $(c\bar d)$ systems are both $35.1\degree$, very close to $35.3\degree$ predicted in heavy-quark limit. So, for $(c\bar u)$ and $(c\bar d)$ systems, physical state $|1P_l\rangle$  is the $|j_l=\frac{3}{2}\rangle$ dominant narrow state with a small decay constant,  while $|1P_h\rangle$  is the $|j_l=\frac{1}{2}\rangle$ dominant broad state with a large decay constant.
On the other hand, the mixing angle $\theta_{1P}$ for $(c\bar s)$ is $-60.4\degree$, and then $|1P_l\rangle$ corresponds to the $|j_l=\frac{1}{2}\rangle$ dominant broad state $D_{s1}(2460)$  with large decay constant, while the $|1P_h\rangle$  is the $|j_l=\frac{3}{2}\rangle$ dominant narrow  state $D_{s1}(2536)$ with small decay constant. So, without the CCEs, the predicted $D_{s1}(2460) $ would also have a lower mass than $D_{s1}(2536)$, and we have obtained the correct mass order for the $J^P=1^+$ $(c\bar s)$ doublet. The large difference of mixing angles between the $(c\bar u)$ and $(c\bar s)$ systems shows that the light-quark masses may play an important role in the $1^+$ heavy-light mesons.

\begin{figure}[ht]
\centering
\subfigure[$\theta_{1P}$ and $\Delta M$ vs. $m_q$ for $1P (c\bar q)$.]{\includegraphics[width=0.42\textwidth]{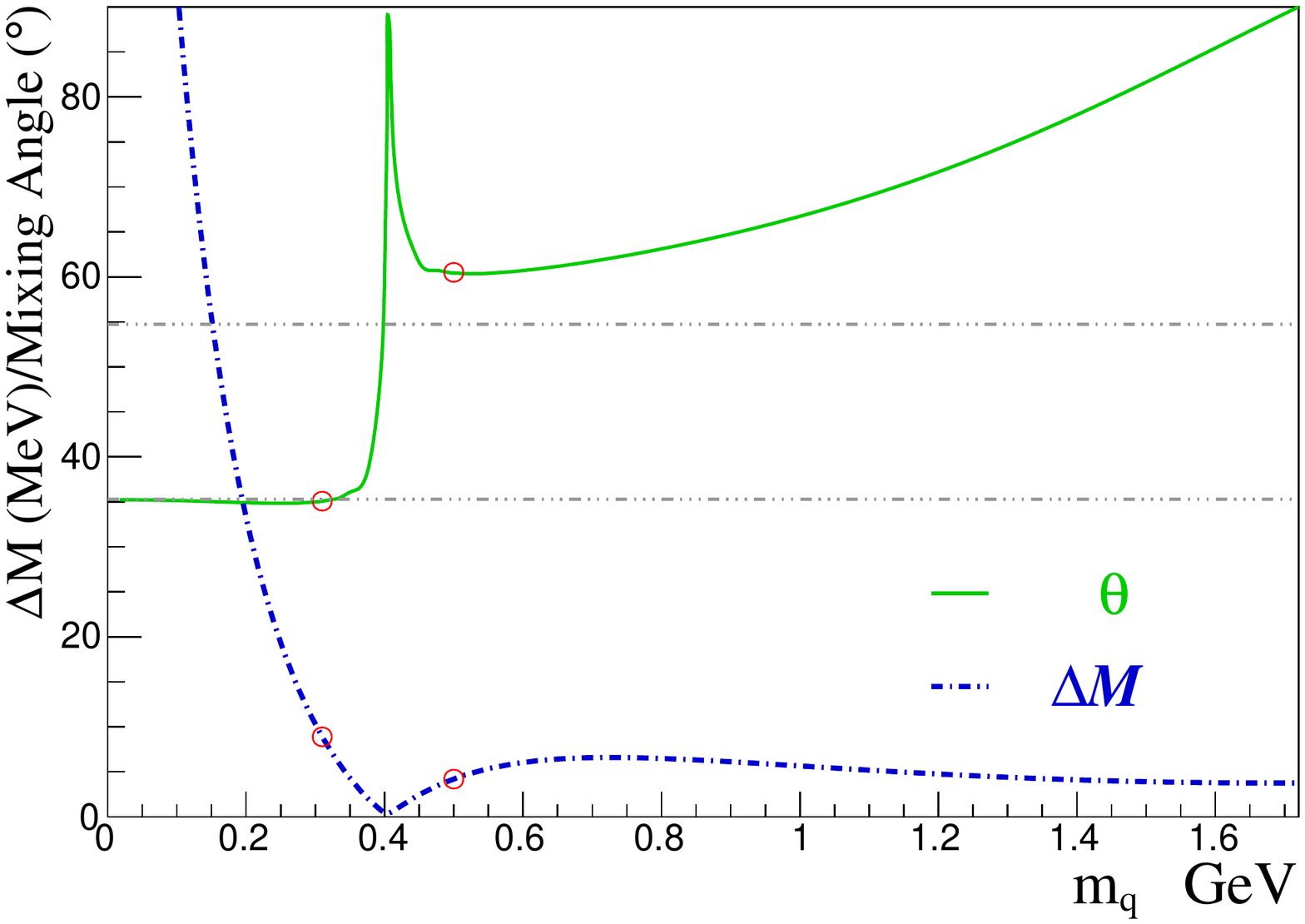} \label{Fig-An-cm2}}
~
\subfigure[Decay constant vs $m_q$ for $1P (c\bar q)$.]{\includegraphics[width=0.42\textwidth]{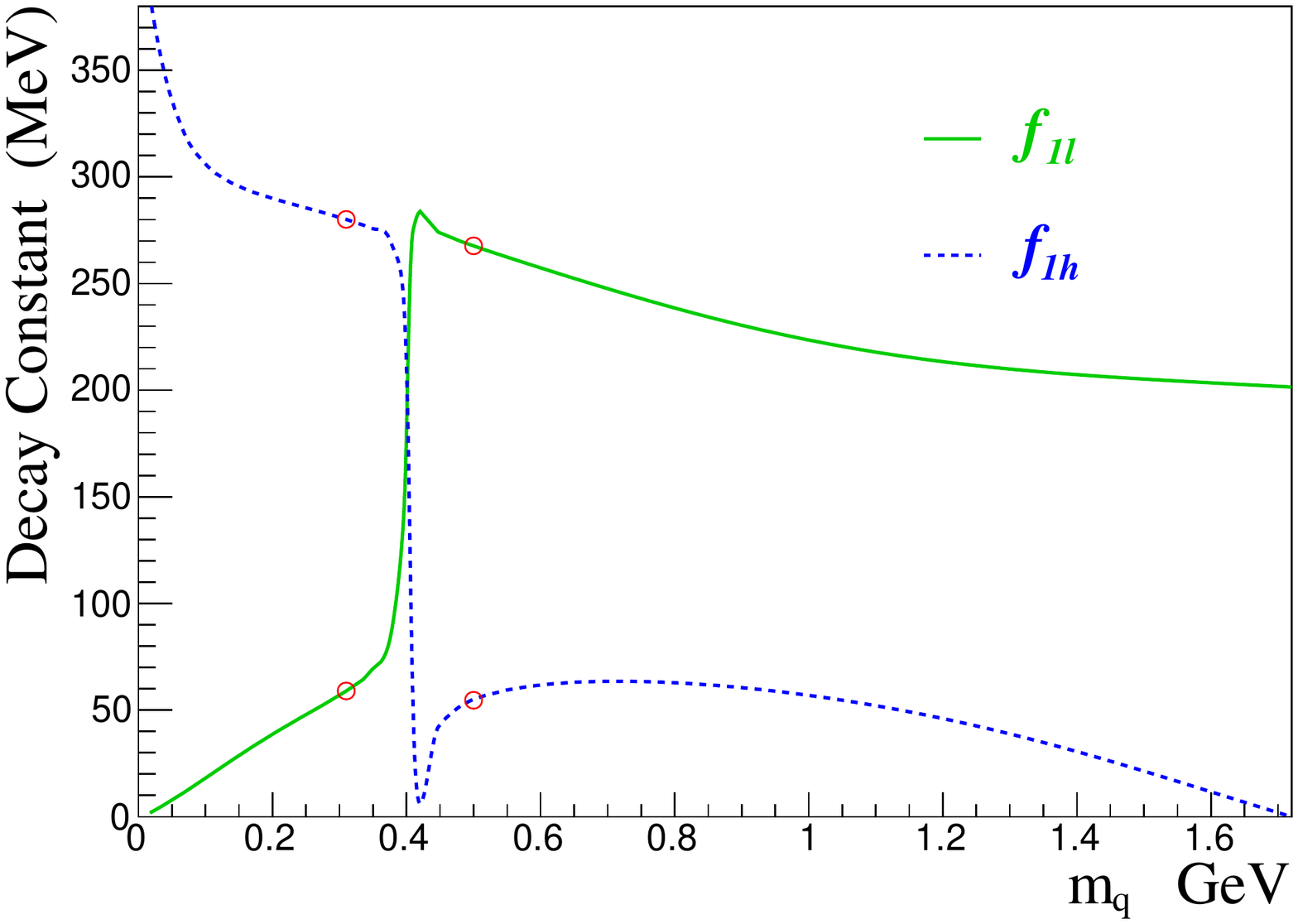} \label{Fig-Dc-cm2}}
\caption{The variation of mixing angle $\theta_{1P}$, mass difference $\Delta M\equiv (M_{1h}-M_{1l})$, and decay constant vs. $m_q$ for $J^P=1^+$ $(c \bar q)$. The circles represent the mass of $m_d$, $m_s$ or $m_c$. A negative sign is added in the mixing angle when $m_q$ is greater than $m_\up{Max}$, where the mixing angle reaches the peak value.}\label{Fig-An-Dc-cq}
\end{figure}

To investigate the relation between light-quark mass $m_q$ and $\theta$ in $J^P=1^+$ $(c\bar q)$ systems, we let $m_q$ change from $0$ to $m_c$ and then explore the mixing angle and $\Delta M\equiv(M_{1h}-M_{1l})$. The obtained numerical results are graphically displayed in \autoref{Fig-An-cm2}. First, when $m_q$ ranges from $0$ to $0.35$ GeV, $\theta_{1P}$ keeps almost constant near the value of $35.3\degree$ predicted in the heavy-quark limit; then increases quickly and reaches the peak at $m_q=m_\up{Max}\simeq 0.4~\si{GeV}$; when $m_q>m_\up{Max}$ the sign of $\theta$ is changed (a negative sign is added in the figure) and the absolute value drops rapidly as $m_q$ increases until about $m_{q}\simeq0.5~\si{GeV}$; finally, $\theta$ increases to $-90\degree$ as $m_q$ closes to $m_c$. On the other hand, the mass difference $\Delta M$ drops rapidly until zero when $m_q$ ranges from 0 to $m_\up{Max}$, and then slowly grows to reach a plateau as $m_q$ increases to $m_c$.

Notice that when $m_q=m_c$, $\theta_{1P}=-90\degree$ means the charmonium system has definite charge conjugation parity, and now the $\ket{1P_l}$ and $\ket{1P_h}$ correspond to the $\chi_{c1}(1P)$ and $h_c(1P)$, respectively. Notice the method is still valid for quarkonia, and the corresponding results here are consistent with what we obtained by solving the $J^{PC}=1^{+-}$ and $1^{++}$ quarkonia directly in Ref.\,\cite{WangGL2007}. The sign of the mixing angle or the mass inversion happens when the light-quark mass is around $0.4~\si{GeV}$, so this inversion picture of the mixing angle can explain well the mass inversion of the $J^P=1^+$ states $D_{s1}(2536) $ and $D_{s1}(2460)$, and partly explain the low mass of $D_{s1}(2460)$.
We also display the dependence of decay constants on $m_q$ for $1^+$ $(c\bar q)$ systems in \autoref{Fig-Dc-cm2}. The variation of decay constants is consistent with the mixing angle.

\begin{figure}[ht]
\centering
\subfigure[$\theta_{1P}$ and $\Delta M$ vs. $m_q$ for $1P (b\bar q)$.]{\includegraphics[width=0.42\textwidth]{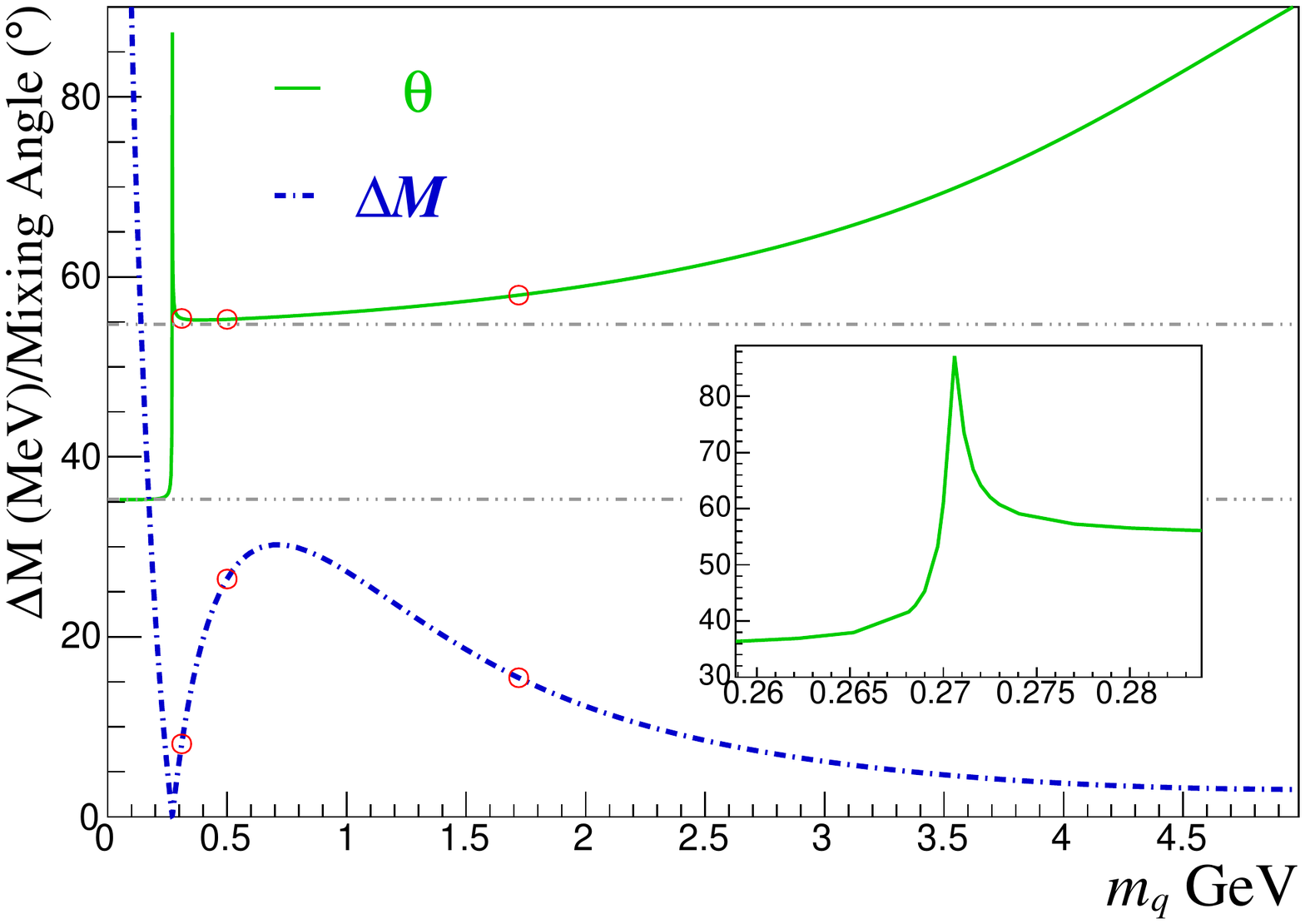} \label{Fig-An-bm2}}
~
\subfigure[Decay constant vs. $m_q$ for $1P (b\bar q)$.]{\includegraphics[width=0.42\textwidth]{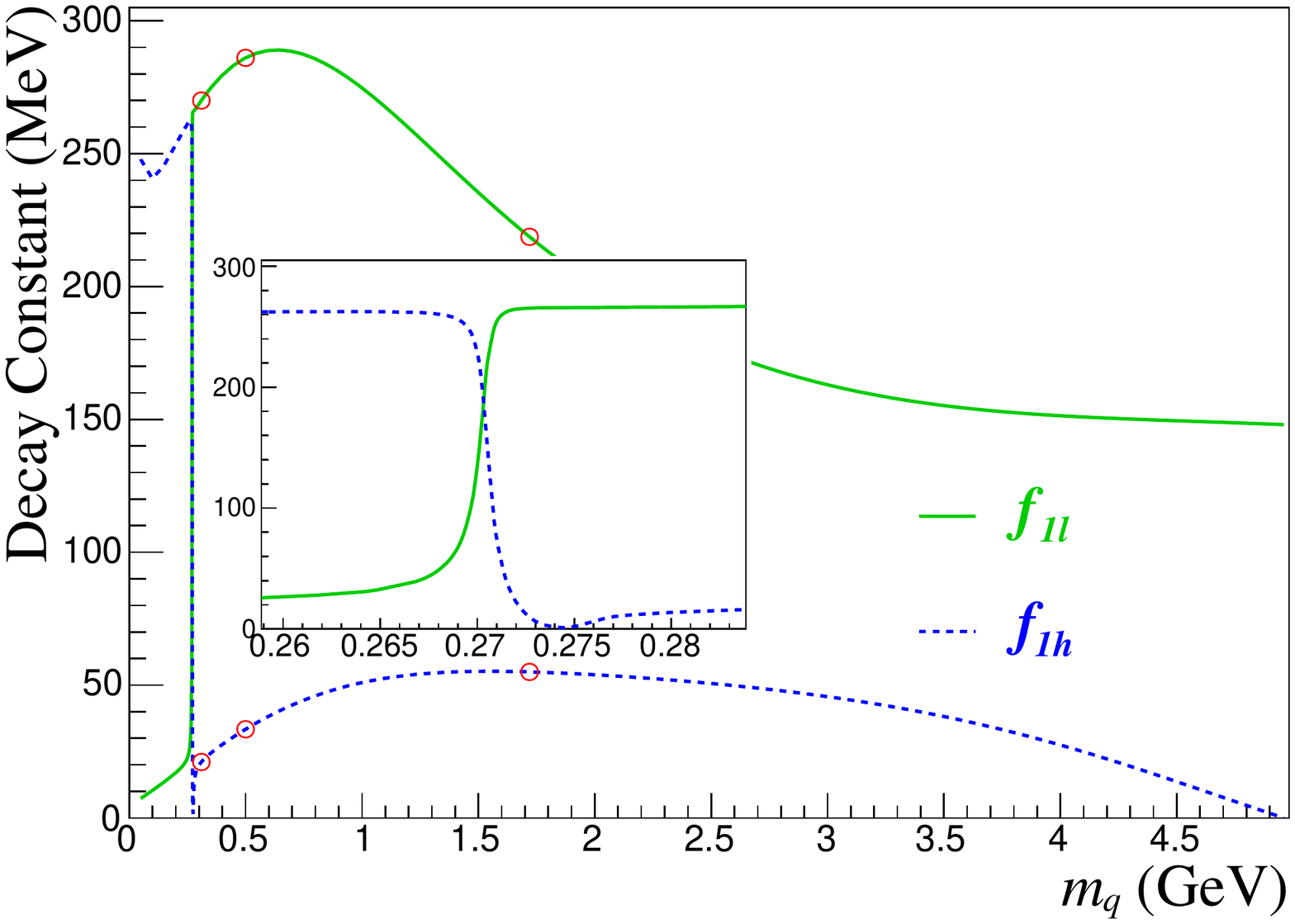} \label{Fig-Dc-bm2}}
\caption{The variation of mixing angle $\theta_{1P}$, mass difference $\Delta M\equiv (M_{1h}-M_{1l})$, and decay constant range along with $m_q$ for $J^P=1^+$ $(b\bar q)$ states. The circles represent the mass of $m_d$, $m_s$ or $m_c$. A negative sign is added in the mixing angle when $m_q$ is greater than $m_\up{Max}$, where the mixing angle reaches the peak value.}\label{Fig-An-Dc-bq}
\end{figure}

The dependence of $\theta_{1P}$ and mass difference $\Delta M\equiv (M_{1h}-M_{1l})$ on $m_q$ for $J^P=1^+$ bottomed states is displayed in \autoref{Fig-An-bm2}, and \autoref{Fig-Dc-bm2} displays the decay constant vs $m_q$.
From \autoref{Tab-M-Dc} and \autoref{Fig-An-bm2}, we can see that, for bottomed $1P$ mesons, the mixing angle inversion happens when $m_q \simeq 0.27$ GeV, which is very close to the constituent masses of $u$- and $d$-quark, but much lower than the $s$-quark mass. So, the mass inversions happen for $(b \bar s)$ and $(b\bar c)$ $1P$ states, while for $(b\bar u)$ and $(b \bar d)$ systems, the inversion phenomenon is sensitive to the choice of light-quark mass. In our calculations, the quark masses $m_u=0.305$ and $m_d=0.311$ GeV are chosen, so the inversions also happen for $(b\bar u)$ and $(b\bar d)$ ground states. The results indicate that, the nonobserved $\ket{\frac{1}{2}}$ dominant broad states $B'_1$ and $B_{s1}'$ are mostly lighter than their partners $B_1(5721)^0$ and $B_{s1}(5830)^0$, respectively. This prediction could also behave as a test on our methods presented here. In Ref.\,\cite{Kalashnikova2016}, the authors also get a similar result within the QCD string model; they obtain $\theta_{1P}=-78.7\degree$ and $B'_1$ is approximately $10~\si{MeV}$ smaller than $B_1(5721)$, which is consistent with our predictions.  For excited states, the situation is different; there is no mass inversion for any of the $2P$ charmed mesons, but for $2P$ bottomed mesons, inversion happens for bottom-stranged and bottom-charmed states.

The decay constants results for $J^P=1^+$ states are listed in \autoref{Tab-Dc} to make a comparison with other studies. Our results of decay constants are close to the previous studies\, \cite{WangGL2007,Veseli1996,Herdoiza2006,ChengHY2004}. From \autoref{Tab-M-Dc}, \autoref{Fig-An-Dc-cq} and \autoref{Fig-An-Dc-bq}, one can see that the decay constant of the narrow $\ket{1P}$ state is usually much smaller than that of its broad $\ket{1P^\prime}$ partner, namely, $f_{1P}\ll f_{1P^\prime}$. Hence the decay constant can behave as a good quantity to distinguish the $J^P=1^+$ doublet of the heavy-light mesons, especially when both states are narrow (because of small phase space, the broad state may have a narrow width) and then hard to be identified by mass and width, such as the situation in $D_{s1}$ and $B_{s1}$ systems.

To see the sensitivity of the results on the model parameters, we calculate the theoretical uncertainties by varying potential parameters $\lambda$, $\Lambda_\text{QCD}$, $a_{1(2)}$, and $V_0$,  and all the constituent quark masses by $\pm3\%$ simultaneously, and then finding the maximum deviation. Considering the uncertainties of parameters, we obtain large ranges of the mixing angle and decay constant for $1^+$ heavy-light states because of the peak structure of special inversion; this may be the reason why a large range mixing angles exists in the literature. We also note that, for $1P$ $(b\bar u)$ and $2P$ $(b\bar s)$ (similar to $b \bar d$ if with larger variation of down-quark mass), the inversion phenomenon is sensitive to the choice of light-quark mass, there may be no inversion within the errors.
\begin{table}[h!]
\setlength{\tabcolsep}{10pt}
  \centering
  \caption{Comparison of the decay constants $f_{1^+}$ for $J^P=1^+$ heavy-light mesons with others' in unit of \si{MeV}. Reference\,\cite{Veseli1996} used the mock-meson approach, Refs.\,\cite{ChengHY2004,Verma2012} used the covariant light-front approach, and Ref.\,\cite{Herdoiza2006} applied the unquenched lattice QCD.}  \label{Tab-Dc}%
    \begin{tabular}{crcccccc}
\toprule[2pt]
    $f_{1^+}$      	&{This} 	&Ref.\,\cite{Veseli1996} 	&Ref.\,\cite{ChengHY2004} &Ref.\,\cite{Verma2012} &Ref.\,\cite{Herdoiza2006} &Ref.\,\cite{WangZ2008} \\
\midrule[1.5pt]
    $f_{D_1}$      	& $56.5\er{5.2}{8.7}$  	&${77\pm18}$ 	& -36   		& -53.6 		& -      		& - \\
    $f_{D_1'}$  	& $266.8\er{8.7}{8.6}$ 	&${251\pm37}$ 	& 130   		& 179   		&{294(88)} 	& - \\
    $f_{D_{s1}}$   	& $54.9\er{47.2}{5.9}$  	&${87\pm19}$ 	& -38   		& -57.3 		& -      		& - \\
    $f_{D_{s1}'}$  	& $267.7\er{9.0}{8.9}$ 	&${233\pm31}$ 	& 122   		& 154   		&{302(39)} 	& - \\
    $f_{B_1}$   	& $21.0\er{13.8}{5.0}$ 	&${32\pm10}$ 	& -15   		& -21.4 		& -      		& - \\
    $f_{B_1'}$  	& $266.6\er{9.0}{8.8}$   	&${206\pm29}$ 	& 140   		& 175   		& -      		& - \\
    $f_{B_{s1}}$   	& $33.4\er{2.4}{2.5}$  	&${36\pm10}$ 	& -      		& -28.3 		& -      		& - \\
    $f_{B_{s1}'}$  	& $286.1\er{7.5}{7.1}$ 	&${196\pm26}$ 	& -     		& 183   		& -      		&${240\pm20}$ \\
    $f_{B_{c1}}$   	& $57.0\er{2.3}{2.3}$  	& -      		& -      		& -47.3 		& -      		& - \\
    $f_{B_{c1}'}$  	& $227.0\er{12.7}{13.4}$ 	& -      		& -      		& 157   		& -      		& - \\
\bottomrule[2pt]
    \end{tabular}%
\end{table}%

\section{Conclusions}\label{Sec-5}

In this work, we have systematically studied the mass spectra, mixing angle and decay constants of the $J^P=1^+$ heavy-light mesons by Bethe-Salpeter methods. For the first time, we obtained the Salpeter wave function of $J^P=1^+$ states without any man-made mixing. Our results indicate that the $1^+$ Salpeter wave function also contains the $S$- and $D$-wave components besides the dominant $P$-wave. We found there is the phenomenon of the mixing angle inversion along with variation of light-quark mass, and this phenomenon results in the mass inversion within the $J^P=1^+$ doublet,  which could explain the mass inversion between $D_{s1}(2536)$ and $D_{s1}(2460)$, and help relieve the low-mass problem of $D_{s1}(2460)$. The mass inversion phenomenon is predicted to exist in the $J^P=1^+$ bottomed mesons. It is worth pointing out that the existence of mass or mixing angle inversion in bottomed system is not sensitive to the choice of the parameters in the potential model but is quite sensitive to the choice of the light-quark mass.
This inversion and peak picture also explained why the obtained mixing angles have confused values with large ranges in the literature. Besides, we also calculated the decay constants and compared our results with others. The decay constants of $|P\rangle$ states are usually much larger than their $|P'\rangle$ partners; this characteristic could provide another quantity to identify the $1^+$ doublet in heavy-light mesons.

\section*{Acknowledgments}
This work was supported in part by the National Natural Science
Foundation of China (NSFC) under Grant Nos.~11575048, 11405037, 11505039, 11447601, 11535002, and 11675239. It was also supported by the China Postdoctoral Science Foundation under Grant No.~2018M641487. We also thank the HPC Studio at Physics Department of Harbin Institute of Technology for access to computing resources through INSPUR-HPC@PHY.HIT.
\medskip

\begin{appendix}
\setcounter{table}{0} 
\renewcommand{\thetable}{A.\Roman{table}}

\section{Decomposition of $J^P=1^+$ Salpeter wave functions} \label{App}
The $J^P=1^+$ Salpeter wave function can be decomposed into two parts according to the properties under charge conjugation transformation, namely, $\varphi_{1^+}=\phi_{1^{+-}}+\phi_{1^{++}}$, where the $\phi_{1^{+\pm}}$ here is not normalized compared with the $\varphi_{1^{+\pm}}$ in \eref{E-phi-l}. Then in terms of the spherical harmonics, $\phi_{1^{+\pm}}$ can be rewritten as
\begin{equation} \label{E-1+-Ylm}
\begin{aligned}
\phi_{1^{+-}} &=C_1(Y_1^{-1} \xi^++  Y_1^{1}\xi^- -Y_1^0\xi^3 ) \left(f_1+f_2 \hat{{\slashed P}} \right)\gamma^5 - C_0 Y_0^0 \frac{\slashed \xi_\perp}{\sqrt{3}} \left(f_3-f_4 \hat{{\slashed P}} \right)\gamma^5  \\
&+C_2\left[Y_2^{-2}\xi^+ \gamma^+ -Y_2^{-1}\frac{ (\xi^3\gamma^+  + \xi^+\gamma^3 )}{\sqrt{2}}  +Y_2^0 \frac{( \slashed \xi_\perp+3 \xi^3\gamma^3)}{\sqrt{6}} - Y_2^1 \frac{(\xi^3\gamma^-+\xi^- \gamma^3 )}{\sqrt{2}} + Y_2^2\xi^- \gamma^- \right] \\
&\times  \left(f_3 - f_4 \hat{{\slashed P}} \right)\gamma^5;
\end{aligned}
\end{equation}
\begin{equation}\label{E-1++Ylm}
\begin{aligned}
\phi_{1^{++}} &=-C_1(Y_1^{-1}\Gamma_\xi^++ Y_1^{1}\Gamma_\xi^-  - Y_1^0\Gamma_\xi^3) \left(h_1 \hat{{\slashed P}} -h_2 \right)\gamma^5 + C_0 Y_0^0 \frac{\slashed \Gamma_{\xi_\perp}}{\sqrt{3}} \left(h_3 \hat{{\slashed P}} +h_4\right)\gamma^5  \\
&-C_2\left[Y_2^{-2}\xi^+ \gamma^+ -Y_2^{-1}\frac{ (\xi^3\gamma^+  + \xi^+\gamma^3 )}{\sqrt{2}}  +Y_2^0 \frac{( \slashed \xi_\perp+3 
\xi^3\gamma^3)}{\sqrt{6}} - Y_2^1 \frac{(\xi^3\gamma^-+\xi^- \gamma^3 )}{\sqrt{2}} + Y_2^2\xi^- \gamma^- \right] \\
&\times  \left(h_3  \hat{{\slashed P}} +h_4\right)\gamma^5,
\end{aligned}
\end{equation}
where $C_1=\sqrt{\frac{4\pi}{3}}$, $C_0=C_1$ and $C_2=\sqrt{\frac{2}{5}}C_1$; $\xi^{\pm} = \mp \frac{1}{\sqrt{2}}(\xi^1\pm i \xi^2)$, $\gamma^{\pm} = \mp\frac{1}{\sqrt{2}} (\gamma^1\pm i \gamma^2)$, $\Gamma_\xi^n \equiv (\xi^n-\slashed \xi \gamma^n)$ with $n=1,2,3$, $\Gamma_\xi^{\pm} = \mp\frac{1}{\sqrt{2}} (\Gamma_\xi^1\pm i \Gamma_\xi^2)$, and $\slashed \Gamma_{\xi_\perp} = (\slashed \xi_\perp -3\slashed \xi)$; $Y_l^m$ is the usual spherical harmonics; $\hat{{\slashed P}}=\frac{\slashed P}{M}$.

From the decomposition \eref{E-1+-Ylm} and \eref{E-1++Ylm} above, considering the relevant coefficients and suppression of $f_{3(4)}$ and $h_{3(4)}$, we can conclude that both $\phi_{1^{+-}}$ and $\phi_{1^{++}}$ contain the $S$, $P$, and $D$-wave components compared with the nonrelativistic description in which only the dominated $P$-wave component is included.
\end{appendix}

\biboptions{numbers,sort&compress}
\setlength{\bibsep}{1ex}  

\end{document}